\documentclass{ws-ijmpa}

\usepackage{graphicx}
\usepackage[super,compress]{cite}
\usepackage{diagbox}
\usepackage{adjustbox}
\usepackage{multirow}
\usepackage{placeins}
\usepackage[T1]{fontenc} 
\usepackage{dsfont}
\usepackage{bm}
\usepackage{color}
\usepackage{placeins}
\usepackage{hyperref}
\hypersetup{colorlinks, allcolors= blue}
\expandafter\let\csname equation*\endcsname\relax
\expandafter\let\csname endequation*\endcsname\relax
\usepackage{mathtools}
\expandafter\let\csname equation*\endcsname\relax 
\expandafter\let\csname endequation*\endcsname\relax 
\usepackage{float,graphicx}
\numberwithin{equation}{subsection}
\usepackage[us,12hr]{datetime} 

\begin{document}

\markboth{T. Lebese and X. Ruan.}{The use of Generative Adversarial Networks to characterise new physics in multi-lepton final states at the LHC}

%
\catchline{}{}{}{}{}
%

\title{\boldmath The use of Generative Adversarial Networks to characterise new physics in multi-lepton final states at the LHC}

\author{Thabang Lebese$^{*}$ and Xifeng Ruan}

\address{School of Physics and Institute for Collider Particle Physics, University of the Witwatersrand,
Johannesburg, Wits 2050, South Africa.\\
tlebese@cern.ch$^{*}$}

\maketitle


\begin{abstract}
Semi-supervision in Machine Learning  can be used in searches for new physics where the signal plus background regions are not labelled. This  strongly reduces model dependency in the search for signals Beyond the Standard Model. This approach displays the drawback in that over-fitting can give rise to fake signals. Tossing toy Monte Carlo (MC) events can be used to estimate the corresponding trials factor through a frequentist inference. However, MC events that are based on full detector simulations are resource intensive. Generative Adversarial Networks (GANs) can be used to mimic MC generators. GANs are powerful generative models, but often suffer from training instability. We henceforth show a review of GANs. We advocate the use of Wasserstein GAN (WGAN) with weight clipping and WGAN with gradient penalty (WGAN-GP) where the norm of gradient of the critic is penalized with respect to its input. Following the emergence of multi-lepton anomalies, we apply GANs for the generation of di-leptons final states in association with $b$-quarks at the LHC. A good agreement between the MC and the WGAN-GP generated events is found for the observables selected in the study.

\end{abstract}

\mbox{}
\clearpage
\section{Introduction} \label{sec:intro}

The discovery of a Higgs boson ($h$) ~\cite{ref-00, ref-01, ref-02, ref-03} at the Large Hadron Collider (LHC) by the
ATLAS~\cite{ref-04} and CMS~\cite{ref-05} experiments has opened a new chapter in particle physics. A new window of opportunity opens for the search of direct evidence of physics beyond the Standard Model (SM) in a field that appears to be.

The use of semi-supervision in Machine Learning (ML) for the search of physics beyond the SM has gained momentum in recent years. ML algorithms are trained on control samples that contain information regarding SM backgrounds and in data samples that may contain new physics in addition of SM backgrounds. Prima facie, this approach does not require a model for the BSM signal. The use of semi-supervision on the basis of topological and signature-based slicing of the phase-space was suggested in Ref.~\cite{Dahbi:2020zjw} to search for these resonances. This approach reduces model dependencies inherent to full supervision, while alleviating poor signal efficiency displayed the  pure semi-supervised approach. 

However, the use of semi-supervision approaches in general is hindered by the possibility of false signals as a result of over-fitting. The probability of false signals needs to be carefully estimated on the basis of toy Monte Carlo (MC) studies. Unfortunately, MC samples based on a full detector simulation is CPU expensive. 


The ATLAS and CMS experiments rely on a Monte Carlo (MC) software for the simulation of events at the LHC as per of the searches for new resonances. Therefore a need to address what can be referred to as the inverse problem in particle physics is of essence, that would be addressing the possibility of whether the extraction of information in order to build a new theory from the data is feasible. For this purpose, particle physicists went beyond the classical methods and introduced machine learning (ML) techniques, a component of artificial intelligence (AI), as an effective aspects for analysing complex and big data, with the hope of eliminating human intervention. With the previously celebrated successes in ML, specifically Deep Neural Networks (DNN), an unsupervised learning technique called Generative-Adversarial Networks (GANs)~\cite{ref0} that can synthesise fake looking samples based on sets of unlabelled training examples. In practice, GANs are mostly used to generate photo-realistic images \cite{ref00}, medical imaging \cite{ref01} and recently with applications in high energy physics for simulating detector responses~\cite{ref02, ref03, ref04}. There are multiple black-box deep learning-based approaches to generative modeling such as Variational Autoencoders (VAE)~\cite{ref-06,ref-07}, Mixture Density Networks (MDN)~\cite{ref-08} and  Generative Adversarial Networks (GAN)~\cite{ref0} that can be used to reproduce the kinematic distributions that are obtained from MC simulations~\cite{ref1}. These models have been considered complimentary to the already work in progress using GANs since they are taken as extensional building blocks to a GAN. 

A number of anomalies have been identified in the production of leptons at the ATLAS and CMS experiments of the LHC~\cite{vonBuddenbrock:2019ajh,vonBuddenbrock:2015ema,vonBuddenbrock:2016rmr,Fang:2017tmh,vonBuddenbrock:2017gvy,vonBuddenbrock:2018xar,Sabatta:2019nfg,Hernandez:2019geu,vonBuddenbrock:2020ter}. These include a number of final states: opposite-sign, same sign di-leptons, three leptons in the presence and absence of $b$-quarks. These final states appear in corners of the phase-space where different SM processes dominate. To date, these anomalies remain unexplained by MC tools that are becoming more and more accurate, while the effects are statistically compelling. Phenomenological analyses indicate that these anomalies can be accommodated by an ansatz composed of two additional scalar fields, $H$ and $S$, where the mass of $H$ is around 270\,GeV and $S$ is SM Higgs-like and has a mass around 150\,GeV~\cite{vonBuddenbrock:2016rmr,vonBuddenbrock:2017gvy,vonBuddenbrock:2019ajh}. In this paper we consider generative modeling to mimic relevant discrimination of observables corresponding to the production of opposite sign leptons in association with $b$-quarks as a show case for this exercise. This approach can be extended for the search of new resonances, as detailed in Ref.~\cite{Dahbi:2020zjw}.

This paper is organised as follows: Section~\ref{sec:lhc} reviews the use of semi-supervision in the study of the multi-lepton problem and the need of GANs for MC studies. Section~\ref{sec:Adv} introduces the GAN techniques, namely the vanilla-GAN, sub-section~\ref{sec:wGAN} is the extension, the Wasserstein GAN (WGAN) and the Wasserstein GAN with gradient penalty (WGAN-GP). Section~\ref{sec:Results} is the results and Section~\ref{Conclusion} concludes the paper.
\section{The multi-lepton problem at the LHC and the role of semi-supervision} \label{sec:lhc}


\subsection{The Multi-lepton problem} \label{sec:Mlep}
In this paper we concentrate on non-resonant searches in signatures with two opposite-sign leptons in association with $b$-quarks. This is response to the multi-lepton anomalies reported in Refs.~\cite{vonBuddenbrock:2019ajh,vonBuddenbrock:2015ema,vonBuddenbrock:2016rmr,Fang:2017tmh,vonBuddenbrock:2017gvy,vonBuddenbrock:2018xar,Sabatta:2019nfg,Hernandez:2019geu,vonBuddenbrock:2020ter}. These anomalies appear to be self-consistent within the framework of a simplified model that introduces two scalars to the SM, $H$ and $S$. In terms of interactions, $H$ is assumed to be linked to electro-weak symmetry breaking in that it has Yukawa couplings and tree-level couplings with the weak vector bosons $V$ ($W^\pm$ and $Z$).
After electro-weak symmetry breaking, the Lagrangian describing $H$ is Higgs boson-like.
Omitting the terms that are irrelevant here, $H$ interacts with the SM particles in the following way:
\begin{equation}
\mathcal{L}_{\text{int}} \supset -\beta_{g}\frac{m_t}{v}t\bar{t}H + \beta_{_V}\frac{m_V^2}{v}g_{\mu\nu}~V^{\mu}V^{\nu}H.  \label{eqn:H_production}
\end{equation}
These are the Higgs-like couplings for $H$ with the top quark, $t$, and the weak vector bosons, respectively.
The strength of each of the couplings is controlled by a free parameter: $\beta_g$ for the $H$-$t$-$t$ interaction and $\beta_V$ for the $H$-$V$-$V$ interaction.
The vacuum expectation value $v$ has a value of approximately $246$\,GeV.
The omitted terms include the Yukawa couplings to the other SM fermions and self-interaction terms for $H$. The first term in Eq.~\ref{eqn:H_production} allows for the gluon fusion production mode of $H$.

In contrast to $H$, the $S$ boson is assumed not to be produced directly but rather through the decay of $H$.
In principle, it is possible to include $S$ as a singlet scalar that has interactions with $H$ and the SM Higgs boson $h$.
Doing this would allow the $H$ to produce $S$ bosons through the $H\to SS$ and $Sh$ decay modes. Studies reported in Ref.~\cite{Hernandez:2019geu} using data from the measurement of the yield of the $Wh$ production mechanism in indicate that BR$(H\to SS) >$  BR$(H\to Sh)$.

These assumptions are all achieved by introducing the following effective interaction Lagrangians.
Firstly, $S$ is given a vacuum expectation value and couples to the scalar sector,
\begin{align}
{\cal L}_{HhS} = &-\frac{1}{2}~v\Big[\lambda_{_{hhS}} hhS + \lambda_{_{hSS}} hSS +
\lambda_{_{HHS}} HHS \notag \\ & + \lambda_{_{HSS}} HSS + \lambda_{_{HhS}} HhS\Big], \label{eqn:HS_coupling}
\end{align}
where the couplings are fixed  to ensure the dominance of the $H\rightarrow Sh,SS$ decays. Secondly, $S$ is given Higgs-like BRs by fixing the parameters in the Lagrangian. For simplicity, the couplings are chosen to be globally re-scaled Higgs-like couplings.

The most copiously produced multi-lepton final state is two opposite sign leptons, due to the decays $S,h\rightarrow W^+W^-$. In this final state the kinematic features of the di-lepton system does not depend significantly on the relative contribution of $(H\to SS)$ and  $(H\to Sh)$ and it is not directly relevant to the studies performed here. As the signal comprises the production of two bosons, the opposite sign di-lepton pair is produced in association with sizeable hadronic jet activity. The signal efficiency in the corner of the phase-space to isolate the non-resonant $W^+W^-$ production (due to the requirement of a full jet veto) is relatively low compared to the phase space used to isolate $t\overline{t}$ production. The latter requires the presence of at least one $b$-quark, where no restrictions are applied on the number of jets. 

Here we concentrate on the production of two opposite sign leptons in association with at least one $b$-quark. The most relevant SM background is the production of $t\overline{t}$ with a contribution from $Wt$ production. Here, GANs are used to mimic the MC predictions for the production of these SM processes using a number of discriminating features that will be described in Section~\ref{dataprep}. It is probably appropriate to note that GANs are not  used here to replace full-fledged MC simulations, but rather to generate selected discriminating features that will be distorted due to the presence of BSM physics. 

\subsection{Machine learning with semi-supervision} \label{sec:supervmethods}

Supervised learning or full supervision has the task to learn a training set made of pairs of points. Let $\mathbf{X}=\left(x_1,...,x_n\right)$ be a set of $n$ points, where $x_i \in \mathbf{X}$ for all $i\in \left[n\right]:={1,...,n}$. The points are pulled from a common distribution on $\mathbf{X}$ and are random variables that are independent and identically distributed. In this setup no prior knowledge of this underlying density is needed nor desired. In supervised learning one needs to define $\mathbf{Y}=\left(y_1,...,y_n\right)$, where $y_i \in \mathbf{Y}$ are referred to as the labels of the examples $x_i$. It is also assumed that the pairs $(x_i,y_i)$ are also randomly sampled over $\mathbf{X}\times \mathbf{Y}$. Full supervision can often be aimed at solving a problem of classification. This is performed through generative or discriminative algorithms. Generative algorithms model how the data is generated, where the conditional probability $\mathds{P}(x|y)$ is inferred. By contrast, discriminative algorithms do not deal with how $x$ is generated, but rather concentrate on modelling $\mathds{P}(y|x)$. Logistic regressions are commonly used to achieve this goal. 

Semi-supervision is a hybrid where there are elements of both supervised and unsupervised learning. Labelled data is provided but not for all data sets. In this setup, the data set $\mathbf{X}=(x_i)_{i\in[n]}$ is spit into two samples. This includes the points $\mathbf{X}_l:=(x_1,..., x_l)$ that are assigned labels $\mathbf{Y}_l:=(x_1,..., x_l)$ and the points $\mathbf{X}_u:=(x_{l+1},..., x_{l+u})$ for which labels are not provided. In the context of searchers for new physics at the LHC, semi-supervision can be applied, where the labelled sample corresponds to the background. The unlabelled sample would contain background in addition to an unknown admixture of signal from new BSM physics. Situations where two more corners of the phase-space with different admixtures of the BSM signal. In any of the cases considered, no prior knowledge of the yield nor the model dependence of the BSM signal are required. In practice, as the task in hand is to efficiently classify between SM backgrounds and new BSM physics, it is essential that a good understanding of background modelling be provided through control samples where BSM signals are expected to be negligible. 

Full supervision through discriminative model building is widely used for searches of new physics in high energy physics (HEP). This approach assumes detailed knowledge of the BSM signal. As a result, optimizations obtained on the basis of full supervision may seriously undercover the exploration of new physics. Searches for bosonic resonances produced in VBF can serve as an illustration of the bias of full supervision. A minor change in the CP structure of the coupling of the boson to weak bosons induces significant changes in the topology of the final state~\cite{Englert:2012xt,Djouadi:2013yb}. When the search is optimised with SM-like couplings, a potential signal can be missed altogether. In order for the search to gain generality, there would be a need for parameters scan of the model to be performed. In the general case, this exercise can become cumbersome and is not performed in practice.

Machine learning can play a significant role to explore a deeper phase-space available at the LHC, where model dependence of the signal needs to be significantly reduced. This can be achieved through the use of semi-supervision resting on the basis of mixing samples. Side-bands, or signal-depleted corners of the phase-space, and signal-enriched samples are defined. These are confronted with each other as two distinct samples, where the ML algorithm performs a classification task. The data in each of the samples is unlabelled, where prior knowledge of the signal modelling is not necessary. The side-band brings insights from the SM background, either in the form of real data or simulated data, depending on the level of realism that the MC displays. 

In order to evaluate the performance of different machine learning approaches, the SM Higgs boson with the di-photon decay was used in Ref.~\cite{Dahbi:2020zjw}. This allows the understanding of the performance of machine learning in a situation where the particle is produced via different production mechanisms, thus populating various corners of the phase-space. The performance on full supervision approach was compared to that of weak supervision and usage of unlabelled data on full supervision. The latter performs full supervision on a signal sample for which the different production mechanisms are not labelled. The performance of these approaches was compared with that of semi-supervision performed with unlabelled signal samples. The performance on semi-supervision is significantly worse compared to that on full supervision for all the production mechanisms considered here. The effect is strongest for the $t\overline{t}h$ mechanism, which is the one that provides the largest amount of distinct signatures. The VBF mechanism, which provides a distinct topology, is the second most affected. This seems to indicate that semi-supervision alone is not particularly efficient in disentangling signal and background in the presence of signatures and topologies. 

In order to alleviate the effects, Ref.~\cite{Dahbi:2020zjw} introduced signatures and looses topological requirements before implementing semi-supervision training. The impact on the signal efficiency and background rejection in this approach was evaluated for different production mechanisms, where significant improvements are observed with respect to the implementation of semi-supervision inclusively. This is referred to as guided semi-supervision. While semi-supervision, as setup here, has the advantage of not relying on a model for the signal it is necessary to constrain the phase-space where the side-band and the signal region are confronted with each other. While signature and topological requirements are driven by physics considerations, the search is not biased by the phenomenology of a model with a particular set of parameters. 

\subsection{Over-fitting with semi-supervision} \label{sec:OvTr}

The problem of over-fitting can be understood purely as a statistical phenomenon of $2$ parts, a manual and randomised ML data splitting strategy. For a manual splitting strategy, one splits the data-set according to their desire. Albeit, this always results into finding a particular case where the data is different due to statistical fluctuations. As an example, what happens in a manual split data split is; if we have one variable as our training data feature $\mathbf{X}=\left(x_1,...,x_n\right)$. If we subdivide $\mathbf{X}$ into $n$ sub-pieces, say $n = 10$. There are $10$ chances that at least one of them has a bump, i.e, a point of interest, like a signal. But if we have $n = 1000$ data sub-pieces, then the chances increase. For randomised ML data splits, we can have 2 or more cases where we fractionally split the data into train-test sub-pieces or split to include a cross validation set resulting to train, validation and test data-sets using either the K-Folds Cross Validation or the Leave One Out Cross Validation (LOOCV) methods. In each of the ML splits, the problem becomes harder since one is never sure if the observed bump is real or fake in each of the splits; or is particularly due to the selection of the phase space i.e., the selection specifically due to the biasness of the ML method. 

\begin{figure}[H] 
    \begin{center}
    \includegraphics[width=\textwidth,clip]{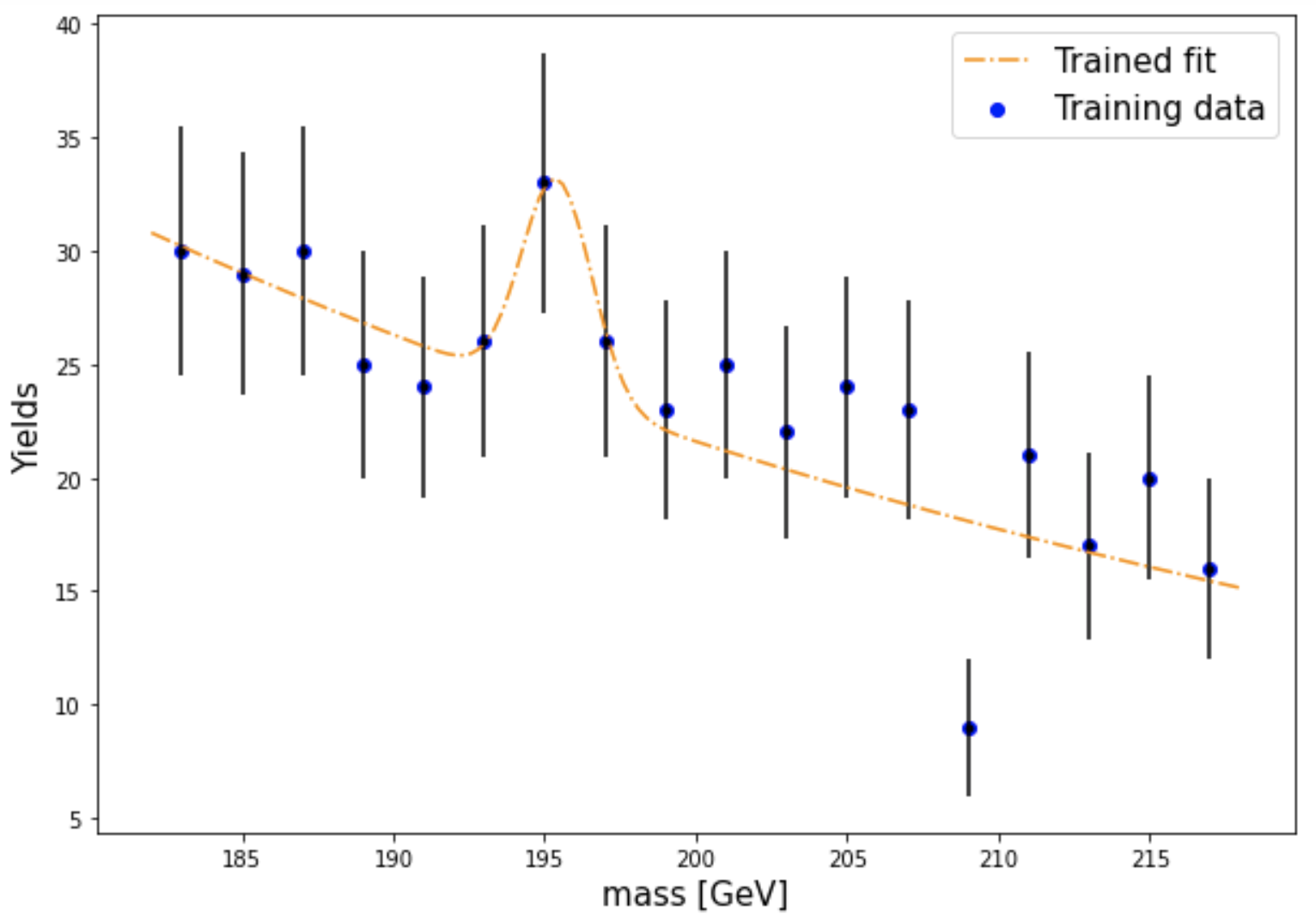}
    \end{center}
\caption{Visualization of fake signal due to over-fitting, with a peak spotted on training data.}
\label{fig:FakeSig}
\end{figure}

With that said, this problem arises in particle physics where there is a concern about the $p$-value deviating from the background-only hypothesis, where it becomes hard to estimate. To estimate the $p$-value, MC simulation is conventionally used to setup the event selection while the real data will be revealed afterwards independently. One of the drawbacks of semi-supervision lies in the fact that over-fitting may engender false signals. This shortcoming becomes particularly acute with the size of the data sample, whether labelled or unlabelled. Though, cross-validation is implemented to overcome over-fitting, the probability that a fake signal exists due to the bias of the training or the statistical fluctuation still needs to be evaluated, dubbed the look-elsewhere effect. As shown in Figure~\ref{fig:FakeSig}, a fake signal peak rises due to over-fitting in a semi-supervision simulation of a background-only case. This issue still remains an unsolved case as it is overlooked in the work by Refs. \cite{Aad:2020cws}. To quantify the probability of such situation, a large background-only MC sample with the same event topology, preferably generated by GAN is necessary. Hence, another potential use case of GANs in this context is to address the inverse problem in particle physics. Where GANs are used to mimic the data in the signal-plus-background regions, essentially becoming generators of new physics.

\section{Review of Adversarial Networks} \label{sec:Adv}

In sub-section \ref{sec:GAN}, we review a vanilla generative adversarial network (GAN) \cite{ref0}, then we consider an improved version by introducing a Wasserstein generative adversarial network (WGAN) in sub-section \ref{sec:wGAN} that does a gradient clipping \cite{ref12}. And lastly we consider the Wasserstein generative adversarial network with gradient penalty (WGAN-GP) that penalizes gradients using a fixed penalty term \cite{ref13}.

\subsection{Generative Adversarial Networks} \label{sec:GAN}

The technique is first introduced by \cite{ref0}, as a class of unsupervised generative models that are deployed in an adversarial settings of two network blocks that implicitly learn the underlying probability distribution of a given data-set. After training, a GAN model provides a mechanism that helps to efficiently sample from the learned distribution with no need for an explicit probability density function. 

From the original inception of GANs \cite{ref0}, a generator network $(G_{\theta})$ and discriminator network $(D_{\phi})$ are pitted against each other over the parameters $\theta$ and $\phi$ respectively. Here the generator takes input samples from a prior distribution $\mathds{P}_{z}(z)$ over some pre-defined latent variable $\mathbf{z}\in \mathds{R}$, where this is usually set as a uniform or normal distribution which it can serve as a source of variation for $G$. $G$ learns a distribution $\mathds{P}_{g}(\mathbf{x})$ from the given training dataset $\mathbf{x} \in \mathds{R}^{n}$ that approximates the real data-generated distribution $\mathds{P}_{d}(\mathbf{x})$ with $m \ll n$. $G(\mathbf{z})$ is therefore defining a mapping from $\mathds{P}_{\mathbf{z}}$ to the data space. $D$ has a task of correctly distinguishing between real samples drawn from $P_{d}$ and synthesized samples produced by $G$. The inputs to $D$ can either be real or synthesized samples at a time, while it outputs is a scalar value between $0$ and $1$ which are a representation of the probability of the inputs coming from $\mathds{P}_{d}$. In this sense, this constitutes a competitive two-player minimax game with players being $D$ and $G$ with the optimal solution being a Nash-equilibrium.  The training objective of a vanilla-GAN is given by

\begin{equation}
\label{eq:eq1}
V_{\textnormal{GAN}} = \min_{\theta} \max_{\phi} \mathds{E}_{\mathbf{x} \sim \mathds{P}_{d}(\mathbf{x})} \big[\log D_{\phi}(\mathbf{x})] + \mathds{E}_{\mathbf{z} \sim \mathds{P}_{z}} \big[\log(1 - D_{\phi}(G_{\theta}(\mathbf{z})))].
\end{equation}
This objective can be optimized by using the stochastic gradient descent (SGD). It can be proved that, when $D$ is fully optimized before updating $G$, minimizing the objective for $G$ is equivalent to minimizing the Jensen-Shannon Divergence (JSD) defined as:

\begin{equation}
\label{eq:eq1a}
JSD(p \parallel g) = \frac{1}{2} \Big[D_{KL}\Big(p \parallel \frac{p+g}{2}\Big) + D_{KL}\Big(g \parallel \frac{p+g}{2} \Big) \Big].
\end{equation}
The $D_{KL}$ is the Kullback-Leibler divergence, a statistical distance often known as relative entropy. $D_{KL}$ has nice properties such as being mean seeking, where reverse $D_{KL}$ is mode seeking. That is:

\begin{equation}
\label{eq:eq1b}
D_{KL}(p \parallel g) = \int_{-\infty}^\infty p(\mathbf{x}) \log \Big(\frac{p(\mathbf{x})}{g(\mathbf{x})} \Big)d\mathbf{x}.
\end{equation}
Hence a JSD quantifies the similarity between $\mathds{P}_{d}(\mathbf{x})$ and $\mathds{P}_{g}(\mathbf{x})$. In a case where $G$ synthesizes samples that are the same as the real data distribution (i.e. $\mathds{P}_{g} = \mathds{P}_{d}$), at that point $D$ becomes maximally confused and thus $D(\mathbf{x}) = D(G(\mathbf{z})) = 0.5$ (Nash equilibrium).

Several number of issues occur during training of a vanilla-GAN that are related to the fact that training happens as a two-player game setup. Some of the most well-known are better explained in the work by the author \cite{weng:2019vka}. A vanilla-GAN is known to be difficult to train, hence a lot of work has been dedicated to understanding, stabilizing and improvement of training \cite{ref09,ref10, ref11}. One of the problems is that the discriminator is often able to perfectly separate $\mathds{P}_{g}$ and $\mathds{P}_{d}$, which pushed the discriminator's error to $0$ and causes a saturation of the JSD. The underlying cause of this is that both $\mathds{P}_{g}$ and $\mathds{P}_{d}$ are generally concentrated on low dimensional manifolds in data space which results in disjoint supports and subsequently leading to vanishing gradients during training.

\subsection{Wasserstein GAN} \label{sec:wGAN}

Challenges of vanilla-GANs from Section \ref{sec:GAN}, lead to proposal of a popular variant called the Wasserstein-GAN (WGAN) \cite{ref12}, studied and applied in many subareas of HEP. A WGAN differs from the vanilla-GAN in that it minimizes the Earth-Mover distance (also known as the Wasserstein distance) $W(\mathds{P}_{d}, \mathds{P}_{g})$ as an alternative distance measure for training a GAN. Informally it can be interpreted as the minimum amount of energy required to transform one probability mass over a distance so as to transform a distribution $\mathds{P}_{g}$ into a target distribution $\mathds{P}_{d}$. Thus:

\begin{equation}
\label{eq:eq-2}
W(\mathds{P}_{d}, \mathds{P}_{g}) = \inf_{\gamma \in \Pi(\mathds{P}_{d}, \mathds{P}_{g})} \mathds{E}_{\mathbf{(x,y)} \sim \gamma} \Big[\lVert{x-y}\rVert \Big],
\end{equation}
where $\gamma$ $\in\Pi(\mathds{P}_{d}, \mathds{P}_{g})$ is the joint distribution with respective marginals distributions $\mathds{P}_{d}, \mathds{P}_{g}$ and $(x, y) \sim \gamma$ being the random variable $(x, y)$ drawn from the distribution $\gamma$. The infimum
in eq. (\ref{eq:eq-2}) is intractable, but  with the concept of the Kantorovich-Rubinstein duality, one can transform $W$ to

\begin{equation}
\label{eq:eq-2a}
W(\mathds{P}_{d}, \mathds{P}_{g}) = \sup_{\lVert{f}\rVert_L \leq 1} \Big[ \mathds{E}_{\mathbf{x} \sim \mathds{P}_{d}} [f(\mathbf{x})] - \mathds{E}_{\mathbf{x} \sim \mathds{P}_{g}} [f(\mathbf{x})] \Big], 
\end{equation}

where the supremum is over unitary-Lipschitz functions $f$. A function $f$ is said to be K-Lipschitz if:

\begin{equation}
\label{eq:eq-2b}
\forall \mathbf{x}, \mathbf{y} \exists K < \infty : |\mathbf{f(x)}-\mathbf{f(y)}| \leq K |\mathbf{x} - \mathbf{y}|.
\end{equation}
This mild metric was  introduced to combat the vanishing gradients that often occur when training a vanilla-GAN. A WGAN was constructed based on this concept with the use of the Kantorovich-Rubinstein duality combined the K-Lipschitz Continuity condition. The value function of the WGAN now becomes: 

\begin{equation}
\label{eq:eq2c}
V_{\textnormal{WGAN}}  = \min_{\theta} \max_{\phi \in \Omega}
\Big[ \mathds{E}_{\mathbf{x} \sim \mathds{P}_{d}(\mathbf{x})} \big[D_{\phi}(\mathbf{x})] - \mathds{E}_{\mathbf{z} \sim \mathds{P}_{\mathbf{z}}} \big[ D_{\phi}(G_{\theta}(\mathbf{z}))] \Big],
\end{equation}
where $\Omega$ is made up of a family of 1-Lipschitz continuous functions (K$=1$). In this formalism $D$ no longer plays a role of a classifier but rather a regressor, hence referred to as a critic. It is tasked to learn a function that approximates $W(\mathds{P}_{g}, \mathds{P}_{d})$. Unlike in a vanilla-GAN, the WGAN has a critic that does not suffer from vanishing gradients. This is a major improvement, since it supports cases where the distributions do not overlap. Therefore, the critic can be trained to optimality by performing multiple training updates on $D$ for every training update on $G$, this can be referred to as a clipping strategy \cite{ref12}.

With the Minimizing Equation (\ref{eq:eq2c}), with respect to the parameter $\theta$ minimizes the learning function $W(\mathds{P}_{g}, \mathds{P}_{d})$. The 1-Lipschitz constraint on the critic is essential since it's for enforcing the clipping the weights of the critic \cite{ref12}. However, \cite{ref13} made a further improvement by proposing an implementation that constrains by adding a gradient
penalty (GP) term to the critic’s objective which penalises the objective whenever the norm of the critic’s gradients exceeds $1$. This further improves the value function by adding a penalty term, making the function to be:

\begin{equation}
\label{eq:eq3}
V_{\textnormal{WGAN-GP}}  = V_{\textnormal{WGAN}} + \lambda \mathds{E}_{\mathbf{\mathbf{\hat x}\sim \mathds{P}_{d}(\mathbf{\hat x})}} \Big[(\lVert{\nabla_{\mathbf{\hat x}}D_{\phi}(\mathbf{\hat x})}\rVert_{2} - 1)^2 \Big],
\end{equation}
with $\mathbf{ \hat x}$ being samples drawn at random locations that lie along a straight line that connects true data distribution $\mathds{P}_{d}$ and the generated distribution $\mathds{P}_{g}$, this is a procedure of sampling from $\mathds{P}_{\mathbf{\tilde x}}$. It is such that $\mathbf{\tilde x} = G_{\theta}(\mathbf{z})$ and $\mathbf{x}$ with $t \in [0,1]$ uniformly sampled, making the straight line to be $\mathbf{\hat x} = t \mathbf{\tilde x} + (1-t) \mathbf{x}$. In this approach, the gradients of the critic are approximated with respect to the inputs and the enforcement of the $1$-Lipschitz constraint along these lines as an attempt to avoid intractibility issues.

\subsection{Related GAN work in HEP } \label{sec:HEPGAN}

The works on GANs became an appealing method within HEP, there are a number of applications ranging from images to non-image data applications. The earlier works within HEP are strictly on images, and that was short lived since that part could not address some critical questions. One of the earliest non-image data GAN within HEP was around the years of 2018 by Ref. \cite{Datta:2018mwd} for an unfolding task. The most dominating work and currently an active area of research is mainly focused on , parton showers \cite{Bothmann:2018trh, Monk:2018zsb, Andreassen:2018apy} and event generation by Refs. ~\cite{ref1, Paganini:2017dwg, Lin:2019htn, Hashemi:2019fkn, DiSipio:2019imz, Carrazza:2019cnt, Butter:2019cae, Gao:2020zvv}. Generative models in HEP are also explored in the modeling of muon interactions with dense targets \cite{SHiP:2019gcl}, phase space integration \cite{Badger:2020uow, Klimek:2018mza, Bendavid:2017zhk, Bothmann:2020ywa, Musella:2018rdi}, event subtraction \cite{Butter:2019eyo} and unweighting \cite{Backes:2020vka}.

\section{MC samples \& GAN training} \label{dataprep}

\subsection{Modelling approach} \label{sec:modApp}

Our approach here is to consider the modeling of HEP data using a WGAN-GP. This is an improvement from our earlier work where we considered a six feature problem using a vanilla-GAN. We fully describe the proposed method, going from data preparation in sub-section \ref{sec:dataprepMC} to the training architecture in sub-section \ref{sec:MC2GAN} and training environment setup.
\vspace{-2 cm}

\subsection{Data preparation and processing} \label{sec:dataprepMC}
In this study we generated collider events, these events are generated using a number of separately executed programs that mimics hard scattering, showering and hadronization processes, namely MadGraph5 \cite{ref18} and Pythia8 \cite{ref19}. The response of the detector is simulated using Delphes3 \cite{ref20} that carries out the fast simulation in order to resemble the ATLAS experiment detector processes. For the reproduction of events of interest, electrons with $E_{T} > 25$ GeV are selected such that  they satisfy the ``Tight’’ likelihood. Muons, are selected with a quality-based identification methodology employed to satisfy the working point with an efficiency of $95\%$ to keep the purity of the samples maximized. The  jets that are used in the analysis have a kinematic selection with characteristics such that $P_{T} > 20$ GeV and $| \eta| < 4.5$  is maintained. For event categorisation is only on jets with $P_{T} > 30$ GeV, which is only for those that are considered for jet counting. The missing transverse momentum $E^{miss}_{T}$ is selected such that they are consistent with the originating $\rho \rho$ collisions. Simulation of these events has generator-level cuts applied to it, these are such that the transverse momentum of the photons are set to be greater than $25$ GeV and the diphoton invariant mass in the ranges of $105 - 160$ GeV. At the end MC generation, the MC data is then locally pre-processed where $10\%$ of the tails is cut with re-bucketing to the last bin. This is another approach of dealing with long tails.
\vspace{-3 cm}
\subsection{WGAN-GP architecture} \label{sec:MC2GAN}

In section \ref{sec:lhc} we argued how the multi-lepton problem can be approached with a WGAN-GP since it is capable of providing a good performance and stability during training. In this regard, this improved GAN needs to be able to produce the particle events in a wide range of energies. This can be achieved by uniformly scaling the MC input events that are passed into the WGAN-GP.

\begin{figure}[H]
    \begin{center}
    \includegraphics[width=\textwidth,clip]{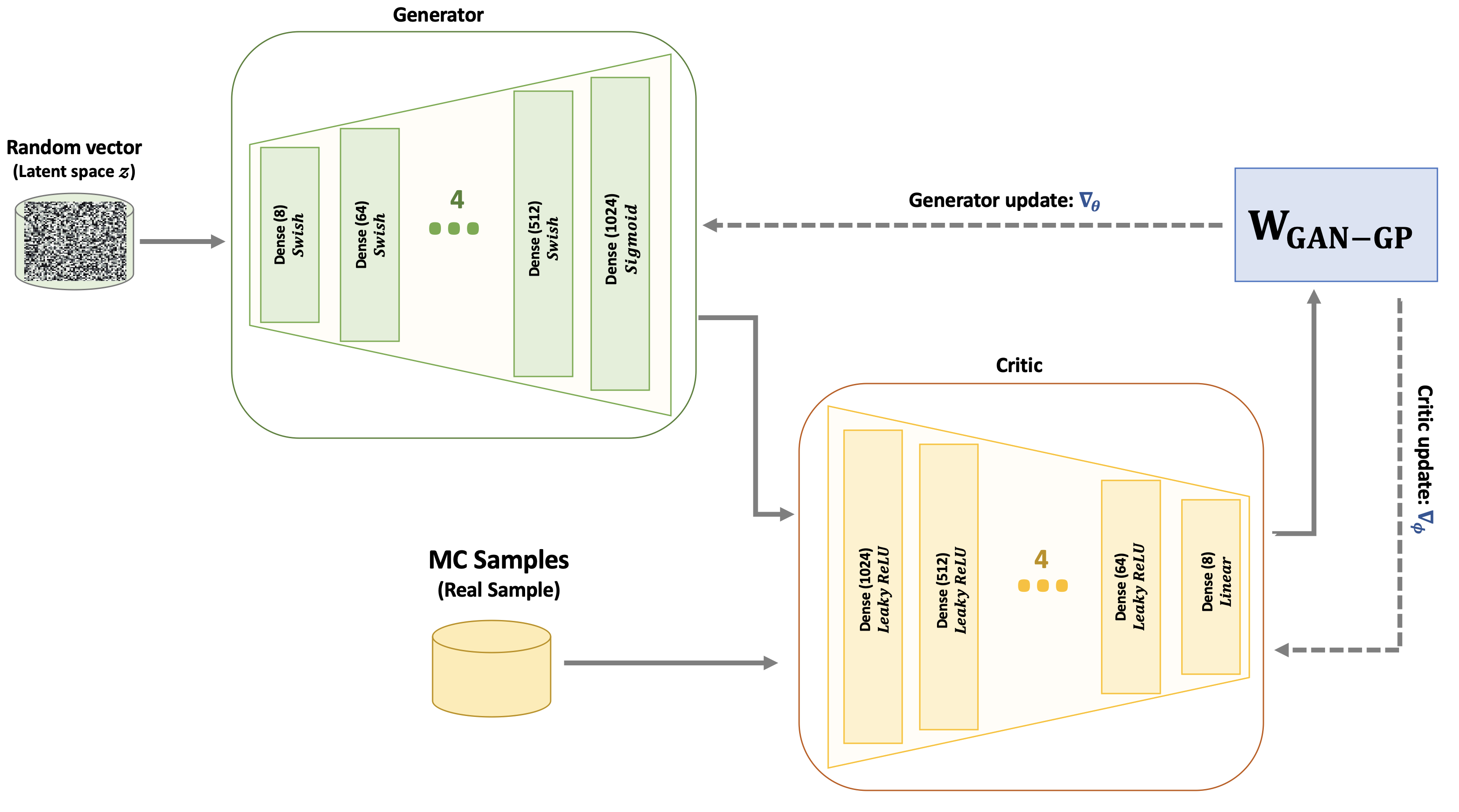}
    \end{center}
\caption{The WGAN-GP Network architecture with a generator (G) and a critic (C). This network is build via a connection of outputs of (G) to inputs of critic (C) whilst penalizing gradient weights with values above $1$.}
\label{fig:gan-gp_pic}
\end{figure}

Given in Figure~\ref{fig:gan-gp_pic} is a descriptive structure of the network topology with the use of MC samples as inputs. Each layer of the WGAN-GP generator (G) has Dense Layers, a trainable $Swish$ function \cite{ref16} as intra-activation functions and a $Sigmoid$ at the end, where (G) consumes the latent space vectors with six hidden layers with in an increasing arrangement culminating to an output layer being the size of the number of MC specific features. A similar-like structure is used in the critic (C) of a WGAN-GP, except that batch normalization is never an option. There are $LeakyReLU$ functions in between and a $Linear$ function used in the final outputs layer(as motivated by \cite{ref12}). The sub-nets of C and G are trained consecutively using a gradient descent and Adam optimizer \cite{ref17} respectively, with a learning rate of $0.0002$ and momentum parameters $\beta_{1} = 0.5, \beta_{3} = 0.999$. However, G is updated with a feedback from C and vice-versa. G is fed a uniformly transformed random noise vectors of the size 128, drawn from Gaussian distributions as inputs in the $[0, 1]$ domain into a vector of 8 features.
The passed features are a representation of the physics kinematics of the leptons in their final state. These are the transverse momentum of the leptons: $p_{\textrm{T}}^{\ell_0}$ and $p_{\textrm{T}}^{\ell_1}$, dilepton transverse momentum and mass: $p_{\textrm{T}}^{\ell\ell}$ and $m_{\ell\ell}$ and
transverse missing energy: $E_{\textrm{T}}^{\textrm{miss}}$, the transverse mass between the transverse missing energy and the dilepton system: $m_{\textrm{T}}$ in $GeV$ and $\Delta\Phi^{\ell\ell}$ in radians and a unit-less $\Delta \textrm{R}^{\ell\ell}$ respectively. These are generated via MC process using the processes in sub-section \ref{sec:dataprepMC}, scale transformed to a range of $[0,1]$ in order to match the samples from the Gaussian. The gradient penalty $\lambda$ is set to $10$, with a training strategy that uses mini-batches of 1024 events for both G and C training in a ($1-$to$-1$) fashion. Since the input vector samples are scaled, this simplifies the learning of the WGAN-GP. Training is carried out for an overall of $1 \times 10^{6}$ epochs, with checkpoint savings at every quarter of the total epochs $(2.5 \times 10^{5})$. This type of a setting enables one to be able to monitor the training process improvements and occasionally eye-ball the performance on the collected results.
\section{Results} \label{sec:Results}

The overall GAN training is carried out on a virtual machine that runs on a Linux environment with a single Tesla P100-PCIE-16GB GPU, using Python programming language, Keras v2.2.4~\cite{ref14} with Tensorflow v1.12 \cite{ref15} back-end as deep learning libraries. The time taken for training is roughly 24 hours for each batch and $4.5$ days in total. 

\begin{figure}[H]
    \begin{center}
    \includegraphics[width=\textwidth,clip]{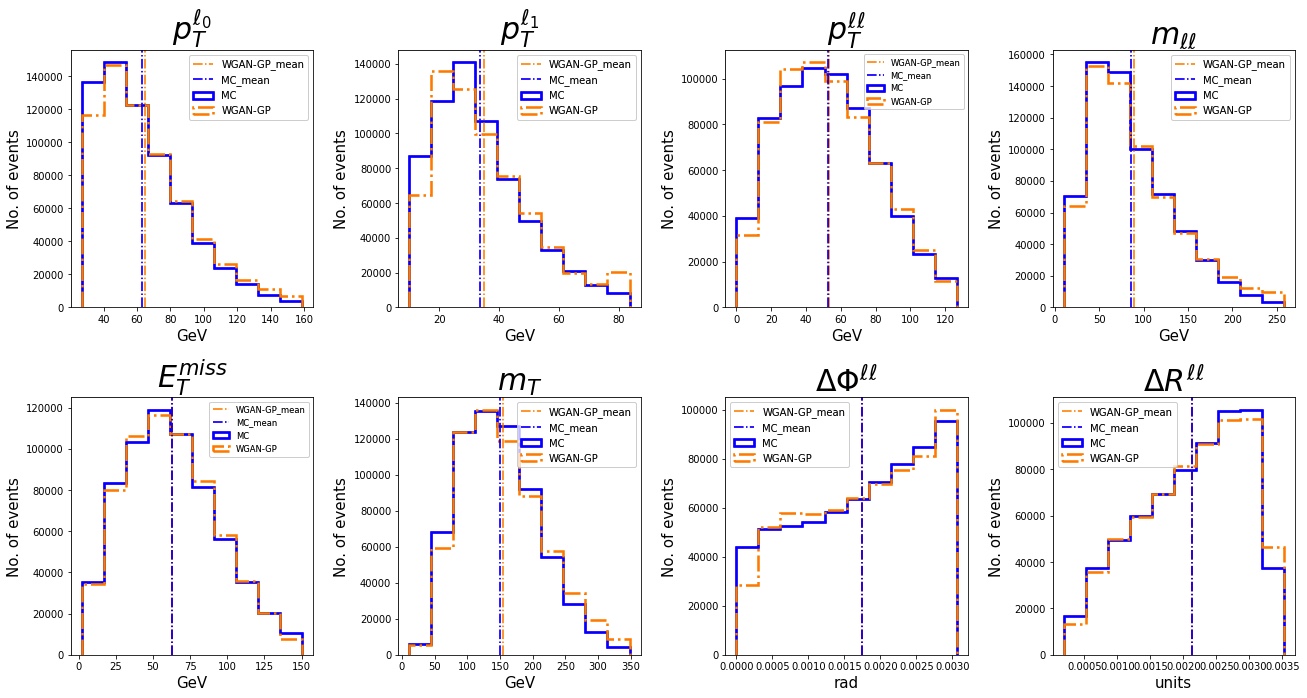}
    \end{center}
\caption{Results on the best performing epoch of the WGAN-GP, with WGAN-GP generated event (in orange) and original MC (in blue) and their associated mean values.}
\label{fig:architecture_scope}
\end{figure}

In the beginning of our work, a number of different generative models were trained and a WGAN-GP showed to be the most consistent in terms of generating better fakes. The WGAN-GP has a better performance, the way it synthesizes events with an increased number of features, i.e., from 6 to 8 features. The individual sample size of each feature increased from about $1 \times 10^{3}$ to $5 \times 10^{5}$ events. We started with a Vanilla-GAN and WGAN with different clipping values $c \in [0.001, 0.1]$. Different hyper-parameters are tried out, including on a WGAN with clipping, the best ones are those used on the WGAN-GP setup. Figure~\ref{fig:architecture_scope} shows the main results of the best epoch performance during training. Each subplot has a horizontal line showing the mean of the MC and GAN. The mean lines give an easy way to observe the uncertainty of generated events during training epochs. When the two mean values coincide, this directly implies a better event generation since it matches the MC mean even though we keep training until each one of the them coincide simultaneously. Table \ref{Tab:models} is the comparison of the quantitative mean values for both MC and WGAN-GP events, their absolute mean differences extracted from Figure~\ref{fig:architecture_scope}. Only one feature shows a relative mean difference of $4\%$, the rest of the feature have a relative difference less than $4 \%$. 

\begin{figure}[H]
    \begin{center}
    \includegraphics[width=\textwidth,clip]{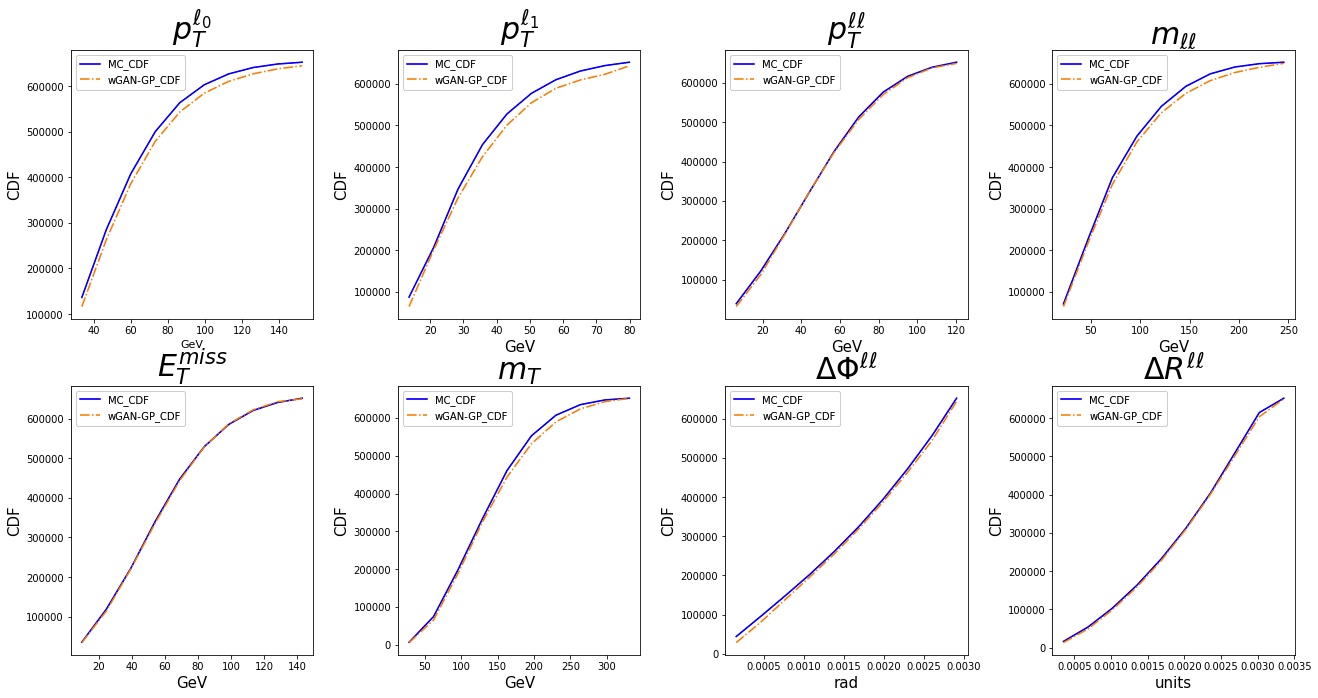}
    \end{center}
\caption{Cumulative Distribution Function (CDF) plot of the MC (in blue) and WGAN-GP (in orange) samples, illustrating the cumulative agreement of the two samples.} 
\label{fig:cum_plot}
\end{figure}
\begin{figure}[H]
    \begin{center}
    \includegraphics[width=\textwidth,clip]{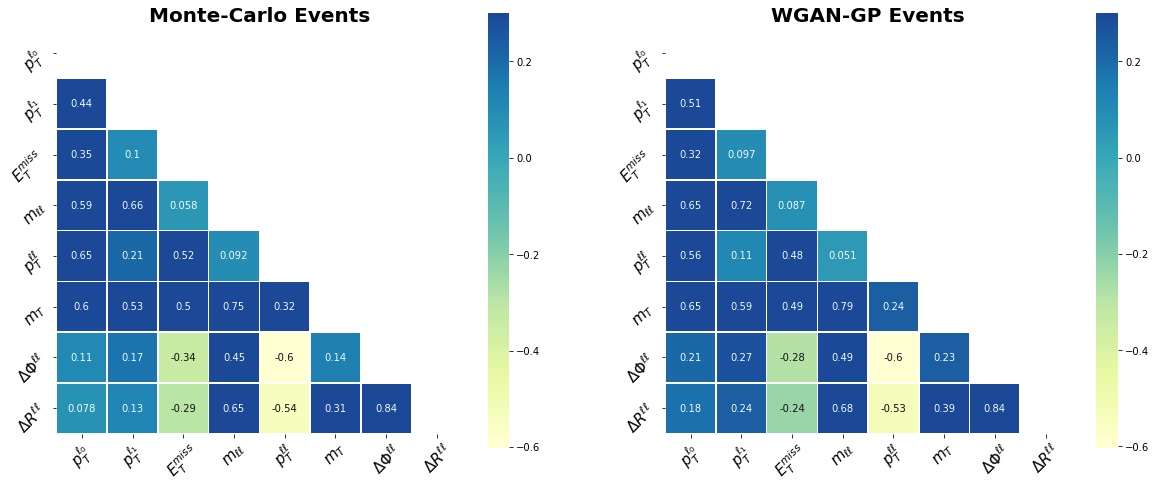}
    \end{center}
\caption{Visualised correlation matrix of the MC and WGAN-GP samples, illustrating the agreement of the two.}
\label{fig:wGAN_corr}
\end{figure}

 Figure~\ref{fig:wGAN_corr} is the visualised correlation matrices of the MC and generated events, this is another way to compare two distributions using the Pearson’s correlation coefficients (PCC) from the two features, i.e., from the MC and GAN in a linear fashion. The Spearman correlation coefficient (SRCC) measures gives the statistical dependence between the rankings of two variables, hence we rely on the PCC measures. Figure~\ref{fig:cum_plot} shows the Cumulative Distribution Function (CDF) plot of results in Figure~\ref{fig:architecture_scope}, there are slight discrepancies for the transverse momentum: $p_{\textrm{T}}^{\ell_0}$  between $80-110$ GeV and  dilepton transverse momentum: $p_{\textrm{T}}^{\ell_1}$ between $40-70$ GeV. The rest of the features show no significant discrepancies.

\begin{table}[H]
\tbl{Tabulated mean values of MC, WGAN-GP events and their relative mean difference.}
{\begin{tabular}{|c|c|c|c|c|} 
\hline
\centering
    Feature $[GeV]$ & MC mean & WGAN-GP mean & Relative difference \\ 
    \hline \hline
    $p_{\textrm{T}}^{\ell_0}$  & 64.876 & 62.987 & 0.030 \\
    $p_{\textrm{T}}^{\ell_1}$ & 35.042 & 33.755 & 0.038 \\
    $p_{\textrm{T}}^{\ell\ell}$ & 53.060 & 52.912 & 0.003 \\
    $m_{\ell\ell}$ & 89.262 & 85.795  & 0.040 \\
    $E_{\textrm{T}}^{\textrm{miss}}$ & 62.480 & 62.543 & 0.001 \\ 
    $m_{\textrm{T}}$ & 154.800 & 149.771 & 0.034 \\
    $\Delta\Phi^{\ell\ell}[rad]$ & 0.002 & 0.002 & 0.004 \\
    $\Delta \textrm{R}^{\ell\ell} [units]$& 0.002 & 0.002 & 0.007 \\
    \hline 
\end{tabular} 
\label{Tab:models}}
\end{table}
\FloatBarrier

\section{Conclusion} \label{Conclusion}
Generative Adversarial Networks are an attractive solution and of potential to searches studies Beyond the Standard Model and to address the inverse problem in HEP. Given that we can synthesize fake MC samples with high accuracy and speed. GANs are an alternative way that allows us to address the multi-lepton problem by simulating the dedicated MC samples and generate large number of events to test the look-elsewhere effect in the semi-supervision study in HEP. GANs work well, but each setup is not easy to generalize for other tasks, each architecture tends to solve one specific problem only. Based on Figures~\ref{fig:architecture_scope}, \ref{fig:cum_plot} and \ref{fig:wGAN_corr} in section \ref{sec:Results}, there is a good agreement between the MC and WGAN-GP generated events. The next step is to further enforce convergence by looking into a conditional WGAN-GP. An attempt on hybrid models such as a combination of Variational Auto-Encoders and Generative Adversarial Networks such as (VAE-GAN) is being carried out. Training time is a major constraint that we facing with the current setup of Keras with Tensorflow back-end, hence it is appropriate that we switch to PyTorch. PyTorch is a perfect alternative that can significantly increase our research productivity whilst scaling on GPUs, making it easy to execute new research ideas. It is capable of decreasing training iteration time on generative modeling from weeks to days.

\section*{Acknowledgements}
The authors want to thank the continued support from the SA-CERN program, hosted by iThemba LABS of the National Research Foundation (NRF) and supported by the South African Department of Science and Innovation (DSI). The authors also want to thank the support from the Research Office of the University of the Witwatersrand.


\end{document}